\documentclass[a4paper,11pt]{article}
\usepackage{pos}
\usepackage{bm}

\title{Entanglement suppression and emergent symmetries in hadron scatterings}

\author*[a]{\mbox{Tao-Ran Hu}}

\author[a]{\mbox{Su Chen}}

\author[b]{\mbox{Katsuyoshi Sone}}

\author[c, a]{\mbox{Feng-Kun Guo}}

\author[b]{\mbox{Tetsuo Hyodo}}

\author[d, e]{\mbox{Ian Low}}

\affiliation[a]{School of Physical Sciences, University of Chinese Academy of Sciences, Beijing 100049, China}

\affiliation[b]{Department of Physics, Tokyo Metropolitan University, Hachioji 192-0397, Japan}

\affiliation[c]{Institute of Theoretical Physics, Chinese Academy of Sciences, Beijing 100190, China}

\affiliation[d]{High Energy Physics Division, Argonne National Laboratory, Argonne, IL 60439, USA}

\affiliation[e]{Department of Physics and Astronomy, Northwestern University, Evanston, IL 60208, USA}

\emailAdd{hutaoran21@mails.ucas.ac.cn}

\abstract{Recently entanglement suppression was proposed to be one potential origin of emergent symmetries. In this work, we extend this theoretical framework to accommodate particles with arbitrary spins and/or arbitrary group representations. As case studies, we discuss recent efforts to test the entanglement-suppression conjecture in two hadron systems that exhibit possible emergent symmetries. The first concerns interactions involving spin-$3/2$ baryons, where entanglement suppression gives rise to symmetries such as SU(40) spin-flavor symmetry. The second system involves low-energy scattering of heavy mesons, where entanglement suppression leads to an enhancement of the inherent heavy-quark spin symmetry to a light-quark spin symmetry, predicting additional siblings for the prominent exotic double-charm meson $T_{cc}(3875)^+$. These predictions should be confronted against experimental data and lattice results to further test the minimal-entanglement conjecture.}

\FullConference{The 21st International Conference on Hadron Spectroscopy and Structure (HADRON2025)\\
 27 - 31 March, 2025\\
Osaka University, Japan\\}

\begin{document}
\maketitle

\section{Introduction}

Emergent symmetries---approximate symmetries absent in the underlying theory---frequently arise in low-energy physical systems, reflecting dynamics and hidden constraints of the relevant degrees of freedom. Recent studies suggest quantum entanglement may play a key role, with entanglement suppression conjectured to enhance symmetry structures~\cite{Beane:2018oxh, Low:2021ufv, Beane:2021zvo, Liu:2022grf, Carena:2023vjc, Liu:2023bnr, Hu:2024hex, Chang:2024wrx, McGinnis:2025brt, Busoni:2025dns, Hu:2025lua}. In quantum information science, entanglement generation of operators, like the $S$-matrix, can be quantified by the entanglement power~\cite{Zanardi:2001zza}. Minimizing this power, called entanglement suppression, acting as a constraint on low-energy hadron reactions, potentially shapes emergent symmetries. Though still conjectural, this approach offers an intriguing framework for understanding symmetry emergence.

The concept of entanglement suppression was first proposed in Ref.~\cite{Beane:2018oxh} as an origin of the emergent spin-flavor symmetry of SU(4) in nuclear forces~\cite{Wigner:1936dx, Wigner:1937zz, Wigner:1939zz}. The emergent SU(2$N_{f}$) spin-flavor symmetry with $N_f$ denoting the number of flavors has been known from the large-$N_{c}$ argument~\cite{Kaplan:1995yg},
but an interesting point is that entanglement suppression can also be achieved by the non-relativistic conformal symmetry~\cite{Birse:1998dk, Mehen:1999nd}. Subsequent studies have explored this topic in various systems, including the pionic scattering~\cite{Beane:2021zvo}, the scattering between light octet baryons~\cite{Liu:2022grf, Liu:2023bnr}, and the perturbative scattering of Higgs doublets~\cite{Carena:2023vjc, Chang:2024wrx, Busoni:2025dns}. It would be interesting to explore consequences of entanglement suppression in more physical systems. 
Furthermore, previous researches have been largely confined to low-spin (or low-isospin) systems; this work aims to establish a general framework for analyzing entanglement suppression in particles with arbitrary spins and/or under arbitrary group representations. As concrete applications, we will then investigate two strongly interacting systems: spin-$3/2$ baryons and heavy mesons.

\section{A general result for entanglement suppression}

In this section, we develop the general framework for analyzing entanglement suppression in the scattering of two distinguishable particles (for the discussion of identical particle scatterings, we refer to Sec. VII in Ref.~\cite{Hu:2025lua}). Our setup considers particles with equal spins (e.g., $np$ scattering), allowing the system in the spin space to be treated as a two-qudit quantum system.

It should be noted that while most literature on entanglement suppression begins by defining operator entanglement powers, this definition can be circumvented entirely. The essence of entanglement suppression lies in treating the two-particle scattering $S$-matrix as a bipartite quantum logic gate, where requiring minimal entanglement power of the $S$-matrix imposes constraints on system parameters~\cite{Beane:2018oxh, Low:2021ufv}. For distinguishable particle scattering, this minimum can always reach zero~\cite{Hu:2025lua}. When the entanglement power of an operator $\hat U$ vanishes, it implies that $\hat U|\psi\rangle$ remains a tensor-product state for any initial state $|\psi\rangle=|\psi_A\rangle\otimes|\psi_B\rangle$~\cite{Hu:2024hex}. However, it has been rigorously proven that for two-qudit systems, only two operators can universally map product states to product states~\cite{10.1063/1.3399808, Johnston01102011, 10.1063/1.3578015, Low:2021ufv}: they are the Identity gate and the SWAP gate.

Identity clearly has a vanishing entanglement power, since it leaves any tensor-product state invariant. The latter, the SWAP operator, is particularly interesting. Being the quantum logic gate exchanging the two qudits,  $\operatorname{SWAP}|\psi_A\rangle\otimes|\psi_B\rangle=|\psi_B\rangle\otimes|\psi_A\rangle$, it still yields a tensor-product state.

Therefore, we can reformulate the entanglement-suppression condition as follows: the $S$-matrix, when acting on a two-qudit quantum system, should be proportional to either the Identity or SWAP gate in order not to generate entanglement. This formulation transforms our problem into determining what constraints must be imposed on the $S$-matrix parameters to satisfy such a requirement. The specific constraints naturally depend on the chosen parameterization of the $S$-matrix. For the simplest case where we consider only spin degrees of freedom, the minimal form of the $S$-matrix would take the structure of
\begin{equation}\label{eq:Sdef}
\hat S=\sum_{J}{\mathcal{J}_{J}\,e^{2i\delta _{J}}},
\end{equation}
where $\mathcal{J}_{J}$'s denote the projection operators onto subspaces belonging to definite irreducible representations (irreps) of the spin group SU(2), with $J$ representing the total spin of this irrep, and $\delta _{J}$'s are the corresponding phase shifts. If one scatters two spin-$s$ particles, the total spin $J$ ranges from 0 to $2s$. Let us then examine how the the Identity or SWAP gate condition constrains the sole parameters in the $S$-matrix, the phase shifts.
\begin{itemize}
\item Identity: Note that the projectors satisfy $\sum_{J}{\mathcal{J}_{J}}=1$, the only solution for making the $S$-matrix proportional to the Identity is to let all phase shifts equal.
\item SWAP: By definition, SWAP interchanges the two particles, thus yielding 1 when acting on symmetric states and $-1$ on antisymmetric states. Thus, it is given by the sum of projectors onto symmetric irreps ($\mathcal{S}_{i}$) minus the sum of projectors onto antisymmetric ones ($\mathcal{A}_{j}$)~\cite{Hu:2025lua}:
\begin{equation}\label{eq:SWAPdef}
\operatorname{SWAP}=\sum_{i}{\mathcal{S}_{i}}-\sum_{j}{\mathcal{A}_{j}}\,.
 \end{equation}
Recall that for half-integral $s$'s, channels with an odd total spin $J$ correspond to the symmetric part of the $s\otimes s$ tensor-product space, while those with an even $J$ correspond to the antisymmetric part (reversed for integral spins). Hence, the SWAP gate takes the form~\cite{Hu:2025lua}:
\begin{equation}
\operatorname{SWAP}=\pm\left(\sum_{\text{odd}~J}\mathcal{J}_{J}-\sum_{\text{even}~J}\mathcal{J}_{J}\right).
\end{equation}
To enforce proportionality of the $S$-matrix to the SWAP operator, the phase shifts must satisfy: $\delta _{J}$'s for all channels with odd $J$ are equal, and similarly, those for even $J$ are also equal, with the two sets differing by $\pi/2$.
\end{itemize}
To sum up, entanglement suppression of Eq.~\eqref{eq:Sdef} is realized, if and only if the phase shifts for odd total spins are all equal ($\delta_\text{odd}$), the phase shifts for even total spins are all equal ($\delta_\text{even}$), and
\begin{equation}
\left|\delta_\text{odd}-\delta_\text{even}\right|=0\,,\quad\text{or}\quad\frac{\pi}{2}\,.
\end{equation}
This is the entanglement-suppression condition for distinguishable particles in the spin space. In the following two sections, we will apply this condition to analyze its physical consequences in two-hadron systems. Our investigation focuses on identifying potential emergent symmetries that may arise from this constraint.

\section{Spin-$3/2$ baryon scattering}

Let us first study the non-relativistic $S$-wave scattering involving the lowest-lying spin-$3/2$ baryons, which can be considered as four-dimensional qudits. These baryons form a ten-dimensional representation $\bm{10}$ (decuplet) under the SU(3) light-flavor symmetry and, in this limit, are considered indistinguishable under strong interactions. This requires a simultaneous treatment of both spin and flavor spaces while incorporating quantum statistics effects. Spin-$3/2$ baryon scattering provides a paradigmatic case for implementing this prescription.

The flavor structure of the decuplet-decuplet system can be decomposed as
\begin{equation}
\bm{10}\otimes\bm{10}=\underbrace{\bm{27}\oplus\bm{28}}_{\text{symmetric}}\oplus\underbrace{\overline{\bm{10}}\oplus\bm{35}}_{\text{antisymmetric}},
\end{equation}
so the corresponding $S$-matrix is given by
\begin{equation}
\hat S=\sum_{J=1,3}\mathcal{J}_{J}\otimes
\left(\sum_{F=\overline{\bm{10}},\bm{35}}\mathcal{F}_{F}\,e^{2i\delta_{JF}}\right)+\sum_{J=0,2}\mathcal{J}_{J}\otimes
\left(\sum_{F=\bm{27},\bm{28}}\mathcal{F}_{F}\,e^{2i\delta_{JF}}\right),
\end{equation}
where we introduce the flavor-space projectors, $\mathcal{F}_{F}$'s, and there are in total eight phase shifts. Due to Fermi-Dirac (FD) statistics, spin-symmetric components ($J=1,3$) project into flavor-antisymmetric irreps, while spin-antisymmetric components ($J=0,2$) project into flavor-symmetric irreps.

As we focus on minimal entanglement in the spin space, it is convenient to define the Identity and SWAP operators acting on the spin and the flavor spaces separately. If the qudits further obey quantum statistics---Bose-Einstein (BE) for bosons or FD for fermions---one can show that~\cite{Liu:2022grf, Hu:2025lua}
\begin{align}
\label{eq:SWAPSWAP}
\operatorname{SWAP}_\text{spin}\otimes\operatorname{SWAP}_\text{flavor}&=\pm1_{\rm spin}\otimes1_{\rm flavor}\,,\\
\label{eq:SWAP1}
\operatorname{SWAP}_\text{spin}\otimes1_\text{flavor}&=\pm1_{\rm spin}\otimes\operatorname{SWAP}_{\rm flavor}\,.
\end{align}
As discussed in the previous section, the non-entangling $S$-matrix in the spin space must be proportional to the Identity or SWAP gate in the spin space. The Identity gate, $\hat S\propto1_{\rm spin}\otimes1_{\rm flavor}$, or the SWAP gate, $\hat S\propto\operatorname{SWAP}_{\rm spin}\otimes1_{\rm flavor}$, can be achieved respectively when~\cite{Hu:2025lua} 
\begin{align}
\delta_{1,\overline{\bm{10}}}=\delta_{3,\overline{\bm{10}}}=\delta_{1,\bm{35}}=\delta_{3,\bm{35}}\,,\quad 
\delta_{0,\bm{27}}=\delta_{2,\bm{27}}=\delta_{0,\bm{28}}=\delta_{2,\bm{28}}\,,\quad 
\left|\delta_{0,\bm{27}}-\delta_{1,\overline{\bm{10}}}\right|=0\text{ or }\frac\pi2\,.
\end{align}
This can be shown using Eqs.~\eqref{eq:SWAPdef},~\eqref{eq:SWAPSWAP}, and~\eqref{eq:SWAP1}. Note that here $\operatorname{SWAP}_{\rm flavor}$ reads as
\begin{equation}
\quad\operatorname{SWAP}_\text{flavor}=\mathcal{F}_{\bm{27}}+\mathcal{F}_{\bm{28}}-\mathcal{F}_{\overline{\bm{10}}}-\mathcal{F}_{\bm{35}}\,.
\end{equation}

Now we can discuss the enhanced symmetries dictated by entanglement suppression. The Identity $S$-matrix requires that all eight phase shifts are equal, corresponding to all channels sharing the same interaction strengths; this leads to an SU(40) spin-flavor symmetry~\cite{Hu:2025lua}. Meanwhile, the SWAP $S$-matrix corresponds to an $\text{SU}(4)_\text{spin}\times\text{SU}(10)_\text{flavor}$ symmetry, as well as the non-relativistic conformal symmetry~\cite{Hu:2025lua}. Note that these emergent symmetries are derived after considering all flavor irreps, though not all scattering sectors contain every irrep. Thus, the above conclusions should be interpreted as consequences of global entanglement-suppression constraints. For sector-specific analyses, see Ref.~\cite{Hu:2025lua}.

\section{Heavy meson scattering}

In this section, we employ the same framework established in the previous section to investigate global entanglement suppression simultaneously in both $D^*D$ and $D^*D^*$ scattering channels.\footnote{The method adopted in this section slightly differs from that in Ref.~\cite{Hu:2024hex}, yet yields consistent conclusions.}

Both $D$ and $D^*$ have isospin $1/2$, and they are pseudoscalar and vector mesons, respectively. Considering BE statistics, we construct the $S$-wave $D^*D^*$ $S$-matrix as
\begin{equation}\label{eq:SD*D*}
\hat S\left(D^*D^*\right)=\mathcal{J} _0\otimes \mathcal{I} _1\,e^{2i\delta _{0*}}+\mathcal{J} _1\otimes \mathcal{I} _0\,e^{2i\delta _{1*}}+\mathcal{J} _2\otimes \mathcal{I} _1\,e^{2i\delta _{2*}},
\end{equation}
where we introduce isospin projectors $\mathcal{I} _I$'s, as well as phase shifts $\delta _{J*}$'s. Its vanishing-entanglement solution is $\delta _{0*}=\delta _{2*}$, $|\delta _{0*}-\delta _{1*}|=0$ (Identity) or $\pi/2$ (SWAP). Additionally, we study the $S$-wave $D^*D$ scattering. Since the total spin is always 1, we focus on isospin entanglement rather than spin entanglement. The corresponding $S$-matrix is given by
\begin{equation}\label{eq:SD*D}
\hat S\left(D^*D\right)=\mathcal{I} _0\,e^{2i\delta _{0}}+\mathcal{I} _1\,e^{2i\delta _{1}},
\end{equation}
with the phase shift denoted as $\delta _{I}$. Its vanishing-entanglement solution is $|\delta _{0}-\delta _{1}|=0$ (Identity) or $\pi/2$ (SWAP).

There are in total five phase shifts in Eqs.~\eqref{eq:SD*D*} and~\eqref{eq:SD*D}, and they are not entirely independent but interrelated through heavy-quark spin symmetry~\cite{Du:2021zzh}. By imposing entanglement-suppression constraints on both $S$-matrices, we will observe how these phase shift relationships become further constrained---effectively enlarging the inherent symmetry.

Moreover, the existence of the isoscalar $T_{cc}(3875)^+$ state~\cite{LHCb:2021auc} so close to the $D^*D$ threshold~\cite{Du:2021zzh} implies that the near-threshold $S$-wave interaction in this $I=0$ channel approaches the unitary limit, so we can further require $\delta_0=\pi/2$ in Eq.~\eqref{eq:SD*D}.

Combining heavy-quark spin symmetry, entanglement suppression, and the additional input, we finally arrive at two possible solutions: (i) all phase shifts are equal to $\pi/2$; (ii) all isoscalar phase shifts are equal to $\pi/2$, and all isovector phase shifts vanish. In both scenarios, the interaction between the heavy mesons does not depend on the total angular momentum, which is referred to as the light-quark spin symmetry, first proposed in Ref.~\cite{Voloshin:2016cgm}. Furthermore, solution (i) would predict isovector hadronic molecules near the $D^{(*)}D^{(*)}$ threshold, as partners of the $T_{cc}(3875)^+$~\cite{Hu:2024hex}.

\section*{Acknowledgments}

TRH, SC and FKG are supported in part by the China National Training Program of Innovation and Entrepreneurship for Undergraduates under Grant No. 202414430004,  by NSFC under Grants No. 12125507, No. 12361141819, and No. 12447101, and by CAS under Grant No. YSBR-101; KS and TH are supported by the Grants-in-Aid  for Scientific Research from JSPSs (Grants No. JP23H05439, and No. JP22K03637); and by JST under  Grant No. JPMJFS2123; IL is supported in part by the U.S.  Department of Energy under contracts DE-AC02-06CH11357, DE-SC0023522, DE-SC0010143 and  No. 89243024CSC000002 (QuantISED Program).


\begin{thebibliography}{99}

\bibitem{Beane:2018oxh}
S. R. Beane, D. B. Kaplan, N. Klco, and M. J. Savage, Phys. Rev. Lett. \textbf{122}, 102001 (2019)
\bibitem{Low:2021ufv}
I. Low and T. Mehen, Phys. Rev. D \textbf{104}, 074014 (2021)
\bibitem{Beane:2021zvo}
S. R. Beane, R. C. Farrell, and M. Varma, Int. J. Mod. Phys. A \textbf{36}, 2150205 (2021)
\bibitem{Liu:2022grf}
Q. Liu, I. Low, and T. Mehen, Phys. Rev. C \textbf{107}, 025204 (2023)
\bibitem{Carena:2023vjc}
M. Carena, I. Low, C. E. M. Wagner, and M.-L. Xiao, Phys. Rev. D \textbf{109}, L051901 (2024)
\bibitem{Liu:2023bnr}
Q. Liu and I. Low, Phys.
Lett. B \textbf{856}, 138899 (2024)
\bibitem{Hu:2024hex}
T.-R. Hu, S. Chen, and F.-K. Guo, Phys. Rev. D \textbf{110}, 014001 (2024)
\bibitem{Chang:2024wrx}
S. Chang and G. Jacobo, Phys. Rev. D \textbf{110}, 096020 (2024)
\bibitem{McGinnis:2025brt}
N. McGinnis, arXiv:2504.21079 [hep-th]
\bibitem{Busoni:2025dns}
G. Busoni, J. Gargalionis, E. Wallace, M. J. White, arXiv:2506.01314 [hep-ph]
\bibitem{Hu:2025lua}
T.-R. Hu, K. Sone, F.-K. Guo, T. Hyodo, and I. Low, arXiv:2506.08960 [hep-ph]
\bibitem{Zanardi:2001zza}
P. Zanardi, Phys. Rev. A \textbf{63}, 040304 (2001)
\bibitem{Wigner:1936dx}
E. P. Wigner, Phys. Rev. \textbf{51}, 106 (1937)
\bibitem{Wigner:1937zz}
E. P. Wigner, Phys. Rev. \textbf{51}, 947 (1937)
\bibitem{Wigner:1939zz}
E. P. Wigner, Phys. Rev. \textbf{56}, 519 (1939)
\bibitem{Kaplan:1995yg}
D. B. Kaplan and M. J. Savage, Phys. Lett. B \textbf{365}, 244 (1996)
\bibitem{Birse:1998dk}
M. C. Birse, J. A. McGovern, and K. G. Richardson, Phys. Lett. B \textbf{464}, 169 (1999)
\bibitem{Mehen:1999nd}
T. Mehen, I. W. Stewart, and M. B. Wise, Phys. Lett. B \textbf{474}, 145 (2000)
\bibitem{10.1063/1.3399808}
E. Alfsen and F. Shultz, J. Math. Phys. \textbf{51}, 052201 (2010)
\bibitem{Johnston01102011}
N. Johnston, Linear Multilinear A. \textbf{59}, 1171 (2011)
\bibitem{10.1063/1.3578015}
S. Friedland, C.-K. Li, Y.-T. Poon, and N.-S. Sze, J. Math. Phys. \textbf{52}, 042203 (2011)
\bibitem{Du:2021zzh}
M.-L. Du, V. Baru, X.-K. Dong, A. Filin, F.-K. Guo, C. Hanhart, A. Nefediev, J. Nieves, and Q. Wang, Phys. Rev. D \textbf{105}, 014024 (2022)
\bibitem{LHCb:2021auc}
R. Aaij \textit{et al}. (LHCb), Nature Commun. \textbf{13}, 3351 (2022)
\bibitem{Voloshin:2016cgm}
M. B. Voloshin, Phys. Rev. D \textbf{93}, 074011 (2016)

\end{thebibliography}
\end{document}